\documentclass[11pt]{article}

\usepackage{graphicx}
\usepackage{color}
\usepackage{amsfonts}
\usepackage{amsmath}
\usepackage{amssymb}

\begin{document}

\begin{titlepage}

\begin{flushright}
%SNS-PH/01-14 \\
\end{flushright}

\vskip 3cm

\begin{center}
{\Large \bf  Gluon fusion production of the Higgs boson \\
in a calculable model with one extra dimension}

\vskip 1.0cm

{\bf
Giacomo Cacciapaglia$^{a,}$\footnote{cacciapa@cibs.sns.it},
Marco Cirelli$^{a,}$\footnote{mcirelli@cibs.sns.it},
Giampaolo Cristadoro$^{b,}$\footnote{cristado@cibs.sns.it}
}

\vskip 0.5cm

$^a$ {\it Scuola Normale Superiore and INFN, Piazza dei Cavalieri 7, 
                 I-56126 Pisa, Italy}\\
$^b$ {\it `Enrico Fermi' Department of Physics, University of Pisa,\\ via Buonarroti 2, I-56126 Pisa, Italy}

\vskip 1.0cm

\abstract{In an extension of the Standard Model with one extra dimension and N=1 supersymmetry compactified on $\mathbb{R}^1/\mathbb{Z}_2 \times \mathbb{Z}'_2$, we compute the Higgs boson decay width into two gluons, relevant to Higgs production in hadronic collisions. 
At one loop, the decay width is significantly suppressed with respect to the SM. 
For a compactification radius $R = (370\pm 70\: GeV)^{-1}$ and a Higgs mass $m_H = 127 \pm 8\: GeV$, as expected in the case of a radiatively generated Fayet-Iliopoulos term, we find it to be less than 15\% of the SM result.}

\end{center}

\vspace{5cm}


\end{titlepage}

\setcounter{footnote}{0}

\section{Introduction}

Recently, a large interest has arisen about the idea that compact extra dimensions can be not so small as previously thought \cite{antoniadis}.
The main drawback, from a phenomenological point of view, is that most of the models based on this idea lack quantitative aspects, so that it is often hard to find a connection with phenomena that take place at a known physical scale.
The model we consider \cite{BHN}, instead, leads to the calculability of several observables, due to the underlying supersymmetric structure of the theory.
In this framework, the Higgs mass \cite{BHN}, the branching ratio $B \rightarrow X_s \gamma$ \cite{BCR} and the muon anomalous magnetic moment \cite{CCC} have already been computed and found insensitive to the UV physics.

Following this line, in this paper we compute the Higgs decay width into two gluons, with the aim of presenting a reliable and cut-off independent calculation in extra dimensions.
The main interest about this process is that it can be related to the cross section for Higgs production \cite{georgi} through gluon fusion.
At a large hadron collider (LHC), this is expected to be the most important channel for light SM Higgs \cite{spira}.
Indeed, in the model we consider, a light Higgs is predicted in the range $m_H \approx 120\div 160\: GeV$.
Moreover, we focus only on zero mode gluon fusion since we expect the parton distribution of KK gluons in the proton to be negligibly small.

\section{The model}

The framework of our calculation is based on the model of ref. \cite{BHN}.
This has the same matter content and gauge structure of the Standard Model, but is embedded in a 5-dimensional space time with N=1 supersymmetry.
The orbifold compactification of the extra dimension on \mbox{$\mathbb{R}^1/\mathbb{Z}_2 \times \mathbb{Z}'_2$} provides supersymmetry breaking {\it \`a la} Scherk-Schwarz \cite{scherk}.
For our purposes the compactification radius $R$ is a parameter and we neglect any dynamical mechanism that stabilizes it.
The presence of only one Higgs hypermultiplet constrains the form of the scalar potential.
As pointed out in ref. \cite{NG-FI}, a Fayet-Iliopoulos (FI) term $\xi D$ can be located on the two branes, thus introducing a free parameter in the potential.
Nevertheless, after ElectroWeak Symmetry Breaking driven by radiative corrections, the Higgs mass is predicted to lie in the range $120\div 160 \: GeV$, while the value of $1/R$, which is correlated to the Higgs mass, can increase up to about a TeV \cite{BHN-FI}.
In the particular case that the FI term is radiatively generated, with a cut-off at a scale $\approx 5/R$, its effect is small enough and the predictions of ref. \cite{BHN} still hold:

$$
m_{H} = 127 \pm 8 \: GeV
$$
$$
1/R = 370 \pm 70 \: GeV.
$$
In \cite{SSSZ} a hypercharge current anomaly is found for the model under consideration. 
It is localized on the two branes and has the property that the integrated anomaly vanishes.
Gauge invariance can however be recovered if the theory is modified in a suitable way.
The overall consistency of the model with the addition of a Chern-Simons term, for instance, is currently under examination.
Anyhow, we believe that our calculation should not be affected by the fixing mechanism.
At one loop, indeed, the Higgs decay is only determined by the brane interactions and does not depend on the (bulk) gauge electroweak structure.
For this reason, also, our results can be extended to a broader class of models: for instance, in a model with two Higgs hypermultiplets \cite{BHN2} the same calculation holds after a proper identification of the Higgs field involved in the vertex.\\

The relevant lagrangian for our calculation comes from the quark Yukawa interactions, located on the branes.
We remind that the vertices of gluon emission are the same for all the KK states, with 4-dimensional coupling $g_{s} = g_{s}^{(5)}/\sqrt{2 \pi R}$.
In terms of mass eigenstates, for every flavour $q$ the Lagrangian is:

\begin{multline}
\mathcal{L}_{int} = - \sum_{n=1}^{\infty} \left\{ \: \left( \tilde{m}^{\pm}_{n} (\phi_H)\right)^2  \big(\Phi_{n}^{\dagger \pm} \Phi_{n}^{\pm} + \Phi_{n}^{c\dagger \pm} \Phi^{c \pm}_{n} \big) + \left( m_{n}^{\pm}(\phi_H)\: \psi^{c \pm}_{n} \psi_{n}^{\pm} +h.c. \right) \: \right\} \\ - \big( m_0 (\phi_H) \: \psi^c_0 \psi_0^{} + h.c. \big)
\end{multline}
where $m_0$, $m_n^{\pm}$ and $\tilde{m}_n^{\pm}$ are the field dependent masses for quark and squark towers.
They have the same form for every flavor $q$, even if the mass matrices arising from Yukawa couplings, $y_q$, located on different branes are not the same (see App. \ref{appendix}):

\begin{equation} \label{masses}
\left\{ \begin{array}{rcl}
\displaystyle \tilde{m}^{\pm}_{n} & = & \frac{2n-1}{R} \pm m_{q}(\phi_H)\\
\displaystyle m^{\pm}_{n} &=& m_{q}(\phi_H) \pm \frac{2n}{R}\\
\displaystyle m_{0} &=& m_{q} (\phi_H) = \frac{2}{\pi R} \mbox{arctan}\left( \frac{\pi}{2} y_q R \phi_H\right)
\end{array} \right.
\end{equation}
Substituting for the complex field $\phi_H=v+H/\sqrt{2}$, this provides us with the full expression of the couplings of the Higgs field $H$ to the mass eigenstates.
Taking the linear terms in $H$ and keeping all orders in $m_q R = \epsilon_q$, we find:

\begin{equation}
\mathcal{L}_{int} = - \lambda_{q}\: \psi^{}_{0} \psi^{c}_{0} H - \lambda_{q} \sum_{n=1}^{\infty} \psi_{n}^{\pm} \psi_{n}^{\pm c} H + h.c. - \lambda_{q} \sum_{n=1}^{\infty}\: (\pm) 2 \tilde{m}_{n}^{\pm}(v) \: \big( \Phi_{n}^{\pm \dagger} \Phi_{n}^{\pm} + \Phi_{n}^{c\pm \dagger} \Phi_{n}^{c\pm}\big) H
\end{equation}
where
\begin{equation}
\label{couplings}
\lambda_q = \frac{1}{\sqrt{2}} y_q \frac{1}{1+\mbox{tan}^2(\frac{\pi}{2} \epsilon_q)} = \sqrt{\frac{G_F}{\sqrt{2}}} \: \frac{2 \sqrt{2}}{\pi R} \frac{\mbox{tan}(\frac{\pi}{2} \epsilon_q)}{1+\mbox{tan}^2(\frac{\pi}{2} \epsilon_q)}
\end{equation}
Note that the Yukawa coupling is related to the Fermi constant $\frac{G_F}{\sqrt{2}} = \frac{1}{4 v^2}$ in a non trivial way, due to the mixing with the massive KK states.
The standard expression, $\lambda_q = \sqrt{\frac{G_F}{\sqrt{2}}} \sqrt{2} m_q$, is recovered for $R\rightarrow 0$.

\section{The calculation}

\begin{figure}[tb]
\begin{center}
\includegraphics[width=12cm]{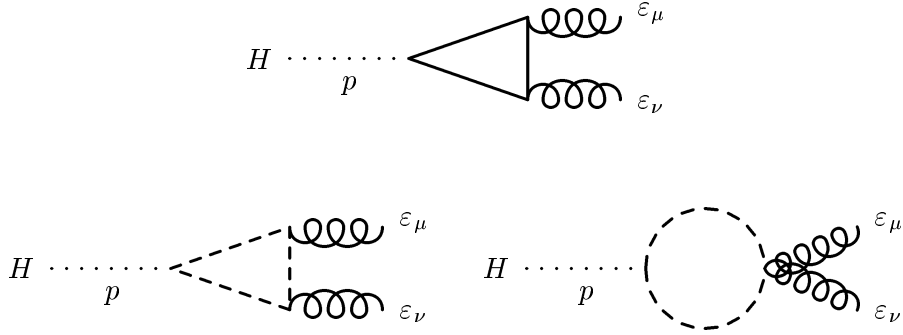}
\parbox{15cm}{\caption{{\small Feynman diagrams contributing at 1-loop to $H \rightarrow gg$. Dashed lines stand for scalars, full lines for fermions.}} \label{grafici}}
\end{center}
\end{figure}

The graphs that contribute at one loop are shown in figure \ref{grafici}.
For a color triplet scalar $\phi$ and a color triplet fermion $\Psi$ with mass $m_s$ and $m_f$ respectively, coupled to the Higgs field via $\mathcal{L}=\lambda_{f} \overline{\Psi} \Psi H + \lambda_s \phi^{\dagger} \phi H$, the amplitude can be written as: 

\begin{equation}
\mathcal{A}^{ab} = - \frac{\alpha_{s}}{4 \pi} \frac{\delta^{ab}}{2} [(p_1 \cdot \varepsilon_2)(p_2 \cdot \varepsilon_1) - (p_1 \cdot p_2)(\varepsilon_1 \cdot \varepsilon_2)]\: \mathcal{F}
\end{equation}
where $\varepsilon_{i}$ and $p_i$ are the gluon polarizations and momenta, and
\begin{equation} \begin{array}{rcl}
\mathcal{F}_{ferm} &=& - 8 \displaystyle \lambda_f m_{f} \int_{0}^{1} dx \int_{0}^{1-x} dy\: \frac{1-4xy}{m_{f}^2 - p^2 xy} \\
\mathcal{F}_{scal} &=& -2 \lambda_s \displaystyle \int_{0}^{1} dx \int_{0}^{1-x} dy\: \frac{4xy}{m_{s}^2 - p^2 x y} .
\end{array} \end{equation}

Let us first review the Standard Model result \cite{SMresult}.
In the present context it comes from the zero mode fermionic graph for $\lambda_{f} = y_q$, and it amounts to:

\begin{equation}
\mathcal{F}_{SM} = \frac{8}{3} \frac{1}{v} \sum_{q} A_{SM}(\xi_q)
\end{equation}
where $\displaystyle \xi_q = \frac{p^2}{m^2_q} = \frac{m^2_H}{m_q^2}$,
$$
A_{SM}(\xi) = \frac{3}{\xi^2} \big\{ 2\xi + \big(\xi-4\big)\: g(\xi) \big\}
$$
and the function $g$ is
$$
g(\xi) = \left\{ \begin{array}{cl}
2 \mbox{arcsin}^2\left(\frac{\sqrt{\xi}}{2}\right) & \xi \leq 4\\
- \frac{1}{2} \left[ \mbox{log}\left( \frac{1+\sqrt{1-4 \xi^{-1}}}{1-\sqrt{1-4 \xi^{-1}}}\right) - i \pi\right]^2 & \xi > 4
\end{array} \right.
$$
The decay width is related to the amplitude, after summing over polarizations and color indices of the final state gluons.
In the SM it is:

\begin{equation}
\Gamma_{SM} (H \rightarrow gg) = \frac{G_{F}}{\sqrt{2}} \frac{\alpha_{s}^2 m_{H}^3}{36 \pi^2} \: \left| \sum_{q} A_{SM} (\xi_q) \right|^2 = \Gamma_0 \: \left|\sum_q A_{SM}(\xi_q) \right|^2
\end{equation}

The main contribution comes from top loops, while the other quarks contribute to a few percent.
The dependence on the Higgs mass, apart from the $m_H^3$ factor in $\Gamma_0$, is encoded in $\xi$, in such a way that the functions $A_{SM}$ run from $\approx$ 0.9, for $m_H=120 \: GeV$, up to $\approx$ 1 for $m_H = 160 \: GeV$.
\\

Let us now move to our model. The couplings are given in eq. (\ref{couplings}), and we consider contributions from top and stop KK towers.
Defining $\epsilon = m_t R$ and $\xi = \displaystyle \frac{m_H^2}{m_t^2}$, the result can be written as:

\begin{eqnarray}
\mathcal{F}_{f} &=& 8 \lambda_{t} R \int dx dy \: (1-4xy) \left\{ \frac{1}{\epsilon} \frac{1}{1-\xi x y} + f(x, y, \xi, \epsilon)\right\} \label{fferm}\\
\mathcal{F}_s &=& 4 \lambda_{t} R \int dx dy \: 4 x y \: s(x, y, \xi, \epsilon)  \times 2 \label{fscal}
\end{eqnarray}
where the factor two in $\mathcal{F}_s$ takes in account the two KK towers of scalar partners and the functions $f$ and $s$ contain the contributions of the KK modes:

\begin{equation} \label{fs}
\left\{ \begin{array}{rcl}
f &=& \displaystyle \sum_{n=1}^{\infty} \left( \frac{\epsilon + 2n}{(2n+\epsilon)^2-\xi \epsilon^2 xy} + \frac{\epsilon - 2n}{(2n-\epsilon)^2-\xi \epsilon^2 xy} \right)\\
s &=& \displaystyle\sum_{n=1}^{\infty} \left( \frac{\epsilon + 2n -1}{(2n-1+\epsilon)^2-\xi \epsilon^2 xy} + \frac{\epsilon - 2n+1}{(2n-1-\epsilon)^2-\xi \epsilon^2 xy}\right)
\end{array} \right.
\end{equation}
Let us remind that the Higgs mass depends on $R$, so that in our model the two parameters $\epsilon$ and $\xi$ are related to each other.

\subsection{$\xi \rightarrow 0$ limit}

To compute the KK state contribution exactly is not a trivial task.
As a first step we note that the sums in $f$ and $s$ simplify in the limit $\xi \rightarrow 0$.
In this case we are neglecting contributions of relative order $\mathcal{O} (m_{H} R)^2$.
These contributions may be important expecially for large $R$, where $m_H \approx 120\: GeV$ and then $m_H R \gtrsim 0.4$. 
However, we use this limit just to have a reliable estimate for small radius ($1/R \approx 1\: TeV$), where $m_H \approx 160 \: GeV$ and then  $m_H R \approx 0.15$, and to have an analytic control on the calculation.

\begin{figure}[ht] 
\begin{center}
\includegraphics[width=10cm]{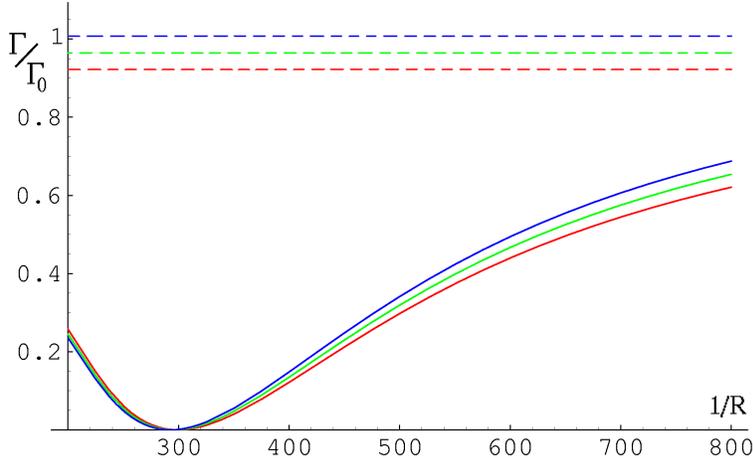}
\parbox{15cm}{\caption{{\small Plot of $\Gamma/\Gamma_0$ as a function of $1/R$, for fixed values of $m_H$ = 120 (\textcolor{red}{red}), 140 (\textcolor{green}{green}) and 160  GeV (\textcolor{blue}{blue}); the upper three lines are the corresponding SM results.}} \label{esatto}}
\end{center}
\end{figure}

In this case, the result is:

\begin{equation} \label{exact}
\Gamma (H\rightarrow gg) = \Gamma_0 \left| \frac{2}{\pi \epsilon} \frac{\mbox{tan}\left(\frac{\pi}{2} \epsilon\right)}{1+\mbox{tan}^2 \left(\frac{\pi}{2} \epsilon\right)} \left[ A_t -1 + \frac{1}{2} \pi \epsilon \mbox{cot}\left(\frac{\pi}{2} \epsilon\right) - \frac{1}{4} \pi \epsilon \mbox{tan}\left(\frac{\pi}{2} \epsilon \right) \right] + A_b \right|^2
\end{equation}
where $A_{t,b} = A_{SM} (\xi_{t,b})$.

As we can see from figure \ref{esatto}, the dependence on the Higgs mass in eq. (\ref{exact}) is smooth in the range $120\div 160\: GeV$ and then the correlation between $m_H$ and $R$ can be consistently neglected.
The decay width is significantly different from the SM one and, in particular, it vanishes for $1/R \approx 300 \: GeV$ (that corresponds to $m_H \approx 120\: GeV$ \cite{BHN-FI}).
This means that the contributions of the top and stop KK modes cancel the SM result.
It is relevant that this happens for every value of $A_b$.\footnote{We have checked that working out a formula similar to eq. (\ref{exact}) including also bottom KK towers, there is no significant change since the parameter $\epsilon_b$ is quite small.}

Also notice that for the unphysical value $1/R = m_t$ the decay width diverges.
This is a remnant of the approximation we are working in, as corrections $\mathcal{O} (m_H R)^2$ are relevant in this region.

\subsection{Full calculation}

In the previous paragraph, the dependence on $m_H$ has been kept only in the SM result. This is, as we saw, an approximation less appropriate for small values of $1/R$.
In order to evaluate how the result fully depends on $m_H$, one could keep the $\xi$ dependence in eq. (\ref{fs}) and then expand in the parameter $\epsilon$.
But, two issues arise.
The first one is that if we expand eq. (\ref{exact}), the series is oscillatory, and stabilizes only at very high order in $\epsilon$.
The second point is that for small $1/R$ the parameter $\epsilon$ is too large and the expansion is not meaningfull.

What one can rather do is to perform the sums in $f$ and $s$ and then evaluate numerically the residual integrals on Feynman parameters $x$ and $y$.
The functions $f$ and $s$ can be simplified to:

\begin{equation}
\begin{array}{rcccl}
f &=& \displaystyle - 2 \epsilon \sum_{n=1}^{\infty} \frac{1}{(2n+\epsilon \sqrt{\xi x y})^2 - \epsilon^2} &=& \frac{1}{2} \left[ \psi \left(\frac{2 + \epsilon \left(\sqrt{\xi x y} - 1\right)}{2}\right) - \psi \left(\frac{2 + \epsilon \left(\sqrt{\xi x y} + 1\right)}{2}\right) \right]\\
s &=& \displaystyle - 2 \epsilon \sum_{n=1}^{\infty} \frac{1}{(2n-1+\epsilon \sqrt{\xi x y})^2 - \epsilon^2} &=& \frac{1}{2} \left[ \psi \left(\frac{1 + \epsilon \left(\sqrt{\xi x y} - 1\right)}{2}\right) - \psi \left(\frac{1 + \epsilon \left(\sqrt{\xi x y} + 1\right)}{2}\right) \right]
\end{array}
\end{equation}
where $\psi (z) = \Gamma'(z)/\Gamma(z)$ is the polygamma function of order zero.
Now, the net result is:

\begin{equation}
\label{gammafin}
\Gamma = \Gamma_0 \left| \frac{2}{\pi \epsilon} \frac{\mbox{tan}\left(\frac{\pi}{2} \epsilon\right)}{1+\mbox{tan}^2 \left(\frac{\pi}{2} \epsilon\right)} \Big[ A_t + 3 \epsilon\: \Xi(\xi, \epsilon) \Big] + A_b \right|^2
\end{equation}
where
$$
\Xi = \int_{0}^{1} dx \int_{0}^{1-x} dy \: \big[ (1-4x y) f + 4 x y s\big].
$$

\begin{figure}[t]
\begin{center}
\includegraphics[width=10cm]{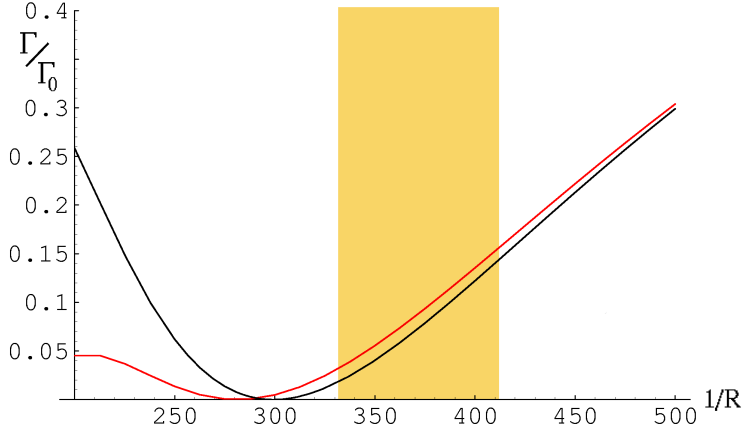}
\parbox{15cm}{\caption{{\small Plot of $\Gamma/\Gamma_0$ as a function of $1/R$ for $m_H = 127 \: GeV$ (\textcolor{red}{red} curve). The black one is eq. (\ref{exact}), i.e. the $\xi \rightarrow 0$ limit.}} \label{xi}}
\end{center}
\end{figure}

Eq. (\ref{gammafin}) contains the exact dependence on the related parameters $\epsilon$ and $\xi$.
In order to estimate how the approximation in the previous paragraph works, we fix $m_H = 127 \: GeV$ and in fig. \ref{xi} we plot the result as a function of $R$ compared to eq. (\ref{exact}) in the same limit.
The two curves differ only slightly in the region $1/R > 300\: GeV$.
Moreover, for small values of $1/R$ the apparent divergence of eq. (\ref{exact}) is not present any more.

In figure \ref{xi}, we can also read the value of the decay width for $1/R = 370\pm 70\: GeV$.
To do this, we first have to take into account the effect of the brane $z$ factors discussed in \cite{BHN}, that are the main source of error on $1/R$.
Eqs. (\ref{exact}) and (\ref{gammafin}) are actually functions of $\tilde{R} = R/(1-\overline{z})$ instead of $R$, since the dependence on the radius only comes from the top and stop mass eigenvalues.
It is $\tilde{R}$ which is determined by the minimization of the Higgs potential, so that its uncertainty does not come from the $z$ factors but from other sources.
It is therefore estimated to be smaller, at 10\% level \cite{BHN-FI}.
Thus, it follows that the decay width, in the case of a radiatively generated FI term, is at 5\% $\div$ 15\% level of the SM one and is only affected at quadratic order in $z$.

\section{Conclusions}

In this paper we have calculated the full one loop amplitude for the decay width of the Higgs boson into two gluons for any value of the compactification radius.
We find a result significantly suppressed with respect to the SM value. 
This is due to a partial cancellation between the SM and the top and stop KK tower contributions. 
In particular, this cancellation is complete and the cross section goes to zero for a value of $1/R \approx 300\: GeV$. 

For the value of the compactification radius of $1/R \sim 370 \pm 70 \: GeV$, as expected in the case of a radiatively generated FI term, the one loop cross section is reduced below the 15\% level of the SM prediction.
Notice, however, that the two loop amplitude, of order $\alpha_s^2$, is expected to be relevant, since in the SM \cite{2loops} it is as large as 40\% of the 1-loop one.

\appendix

\section{Mass matrices} \label{appendix}

In ref. \cite{BCR}, the diagonalization of the mass matrices arising from Yukawa couplings located on the brane $y=0$ is presented.
Dealing with couplings located on the brane $y=\frac{\pi}{2} R$, some subtleties appear.
First of all, the matrices are, for scalars and fermions:

\begin{eqnarray}
\Phi^{\dagger} \mathcal{M}^{2} \Phi & = & \left( \phi^{c \dagger}_{U,l}, \phi^{\dagger}_{T,l} \right) \left( \begin{array}{cc}
\tilde{M}^2 + 4m^2 \eta E  & -2 m E\cdot \tilde{M}\\
-2 m \tilde{M}\cdot E & \tilde{M}^2 \\
\end{array} \right) \left( \begin{array}{c}
\phi_{U,k}^c\\
\phi_{T,k}
\end{array} \right) \\
\overline{\Psi} \mathcal{M} \Psi^{c} &=&  \left( \psi_{T,0}, \psi_{T,l}, \psi^{c}_{U,l} \right) \left(
\begin{array}{ccc}
m & \sqrt{2} m \mathcal{J}^{T} & 0 \\
\sqrt{2} m \mathcal{J} & 2 m \mathcal{J}\mathcal{J}^{T} & M \\
0 & M & 0 \\
\end{array} \right) \left( \begin{array}{c}
\psi_{U,0} \\
\psi_{U,k} \\
\psi^{c}_{T,k}
\end{array} \right)
\end{eqnarray}
where $\eta = \sum_{k} 1$, $M_{lk} = \frac{2l}{R} \delta_{lk}$, $\tilde{M}_{lk} = \frac{2l-1}{R} \delta_{lk}$, $\mathcal{J}_{k} = (-1)^k$, $E_{lk} = (-1)^{l+k}$ and $m=y_q v$.
The difference with ref. \cite{BCR} is the presence of some $-1$ signs that come from the values of the wave functions of the fields, whose dependence on $y$ in given by their parity.
But, carefully performing the calculation, it is clear that the eigenvalues coincide with those produced by the matrices on the brane $y=0$, eq. (\ref{masses}).
Note also that in this case $\phi^c$ is the field in the N=1 supermultiplet of $\psi$.
The effect of these signs is physically relevant only when the two branes interfere.
This happens when the observable receives contributions from both the branes (see for example \cite{BCR}).

The calculation of these matrices deserves another comment.
In fact, if we write the bulk lagrangian as in \cite{arkani} and add the two brane interactions as in \cite{BHN}, it is not trivial how to get rid of the auxiliary fields $F$ and $F^c$.
For $F$ it is straightforward since both the bulk and the $y=0$ brane lagrangians are written in terms of the same multiplets, while for $F^c$ we refer the reader to \cite{MP}.
The result is that the term $\delta (y-\pi R/2) \partial_5 \phi \frac{d}{d \phi^c} W'$, where $W'$ is the superpotential localized on the $y=\pi R/2$ brane, has a minus sign with respect to the analogous term generated by $F$.

\section*{Acknowledgements}

We would like to thank Riccardo Barbieri and Riccardo Rattazzi for useful discussions and suggestions. 
This work is in part supported by the EC grant under the RTN contract HPRN-CT-2000-00148.

\end{document}